# Reduced Graphene Oxide Tattoo as Wearable Proximity Sensor

*Vaishakh Kedambaimoole[1], Neelotpala Kumar[1,3], Vijay Shirhatti[1], Suresh Nuthalapati[1], Saurabh Kumar[2], Mangalore Manjunatha Nayak[2], Prosenjit Sen[2], Deji Akinwande[3], Konandur Rajanna[1,\*]*

[1]Dept of Instrumentation and Applied Physics, Indian Institute of Science, Bangalore 560012, India

[2]Centre for Nano Science and Engineering, Indian Institute of Science, Bangalore 560012, India

[3]Microelectronics Research Center, Department of Electrical and Computer Engineering, The University of Texas at Austin, Austin, TX, USA.

**Corresponding Author:**
K Rajanna, Email: kraj@iisc.ac.in

**Abstract:**

The human body is punctuated with wide array of sensory systems that provide a high evolutionary advantage by facilitating formation of a detailed picture of the immediate surroundings. The sensors range across a wide spectrum, acquiring input from non-contact audio-visual means to contact based input via pressure and temperature. The ambit of sensing can be extended further by imparting the body with increased non-contact sensing capability through the phenomenon of electrostatics. Here we present graphene-based tattoo sensor for proximity sensing, employing the principle of electrostatic gating. The sensor shows a remarkable change in resistance upon exposure to objects surrounded with static charge on them. Compared to prior work in this field, the sensor has demonstrated the highest recorded proximity detection range of 20 cm. It is ultra-thin, highly skin conformal and comes with a facile transfer process such that it can be tattooed on highly curvilinear rough substrates like the human skin, unlike other graphene-based proximity sensors reported before. Present work details the operation of wearable proximity sensor while exploring the effect of mounting body on the working mechanism. A possible role of the sensor as an alerting system against unwarranted contact with objects in public places especially during the current SARS-CoV-2 pandemic has also been explored in the form of an LED bracelet whose color is controlled by the proximity sensor attached to it.

**Introduction:**

Wearable sensors hold the promise of augmenting and aiding the response of human body to both internal and external stimuli. The current pandemic due to Covid19 has highlighted how wearable sensors can be effectively employed to screen, monitor and track the important biomarkers of a human body like pulse, temperature and respiration rate at a mass level and aid in preventing further spread of the infection .[1–3] For any human individual, sensors that acquire knowledge of surrounding are vital to almost all aspects of behavior and cognition. The same chain of thought inspires the architecture of robots in field of automation where their functionality is primarily dependent on external stimuli underpinning the importance of environment sensors. In humans, externally categorized sensors comprise of sensory organs including the eyes, ears, nose, mouth, and skin, corresponding to sense of vision, hearing, smell, taste and touch. These basic five senses can be further divided into contact type and non-contact type defined by their sensing mechanism. The sense of vision, hearing and smell come under non-contact type where the sensing organs acquire information without physical contact with the stimulus while touch and taste are contact type where the sensing is triggered through physical contact with the stimulus.

In humans, sense of touch is the part of the somatosensory system, a cumulative term for the surficial sensing which includes vibration, pressure, temperature and pain, predominantly sensed by the largest organ of the body, the skin. Perception of touch provides a wealth of information to brain like shape and texture of an object besides its temperature and humidity but in a scenario demanding strict social distancing, proximity sensing can bypass the need for contact to detect an object in the ambience without any physical contact. A wearable proximity sensor[4] could serve as an important add-on to the skin, increasing its ambit of sensing beyond the limits set by the make of the human body. Apart from a higher range of sensing, it can be very useful in special situations mirroring current scenario due to Covid-19 pandemic where direct interactions play key role in the spread of the disease. It can be employed as an alerting mechanism to avoid unintentional contact in public places and for keeping track[5] to arrest the disease's outspread.

Besides the obvious benefit of wearable proximity sensors, conventionally, proximity sensors have been utilized in robots employing principles of electromagnetic field or a beam of electromagnetic radiation. Inductive and capacitive type sensors[4,6,7] detect change in electromagnetic field interacting with them with respect to a surrounding object whereas a photosensor detects electromagnetic radiation reflected from the object under observation as a function of its size and distance.[8,9] Traditionally, proximity sensors have been rigid in their construction making them unfit for a comfortable wearable experience. Though there are few studies on flexible proximity sensing,[7,10–12] a graphene-based proximity sensor working on the principle of electrostatic gating that can be tattooed on to the skin has not been reported.

Graphene has remained a highly sought after 2D material since its discovery[13] due to its exceptional electrical[14] and mechanical properties[15–18]. The chemically derived form of graphene known as reduced graphene oxide (rGO)[19] is an inexpensive alternative to CVD grown 2D graphene and is popular among sensor researchers owing to its rich repository of oxygen functional groups[20] which facilitate a multitude of sensing applications.[21–24] The functional groups also influence the electrical properties of graphene which bring about a certain localization of electron depending on its surface chemistry.[25,26] This can be controlled and moderated through appropriate reduction techniques used to reduce graphene oxide (GO). Besides gas sensors[27] and biosensors, various wearable sensors have been fabricated using rGO.[28–30] *Sadasivuni et al* reported flexible proximity sensor based on rGO cellulose nanocomposite[31]. Though this work demonstrated short range detection of human finger, the integration of sensor on skin and the effect of skin on performance of sensor had not been investigated. *Wang et al* have reported an electrostatic field sensor using monolayer graphene deposited on n-doped Si substrate[32]. While the sensor effectively measured proximity of different objects from a larger distance, the device was not flexible to enable integration on the skin since the Si substrate greatly influenced the sensing mechanism.

In the present work, we report a highly flexible and skin conformal proximity sensor fabricated using rGO film that can be tattooed onto the skin. The sensor utilizes the influence of electrostatic charges residing on the surface of the object under observation. Since human skin in general acquires charges on its surface due to friction with atmospheric air, the sensor performance after skin integration was studied extensively as it has not been reported earlier. The working mechanism of the device is neither dependent on the substrate of the sensor nor the protective encapsulation layer thus enabling greater design flexibility in fabricating skin conformal wearable proximity sensors. The cost-effective yet robust sensor with facile fabrication process can be effectively tattooed on the skin and used as a disposable wearable proximity sensor. The application of this sensor can also be extended to robotics and other day to day activities that limit the act of touch during an outbreak of possible pandemic disease.

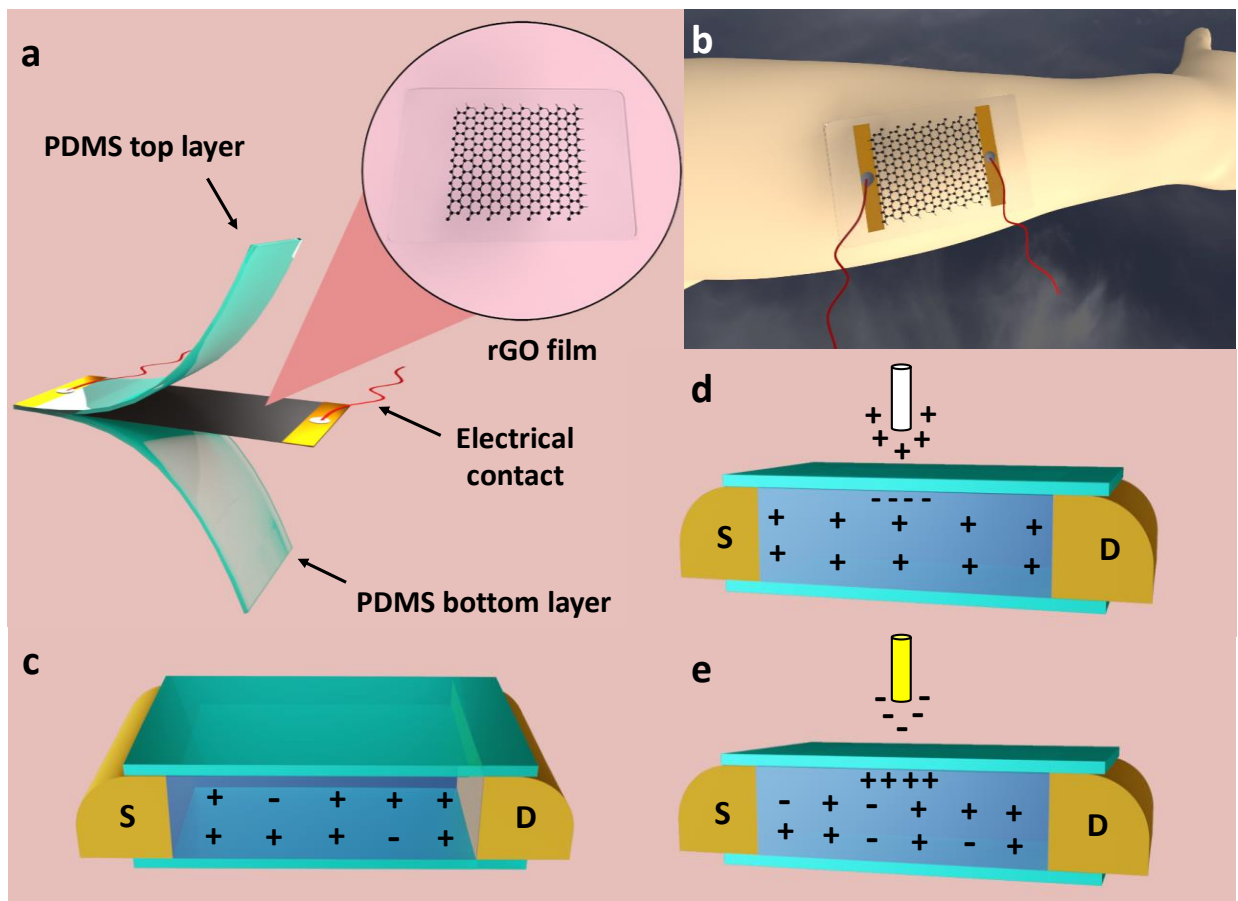

**Figure 1: Schematic and working mechanism of wearable proximity sensor.** a) Components of wearable proximity sensor – rGO thin film encapsulated between ultrathin PDMS layers. b) Sensor tattooed onto human skin for proximity sensing studies. c) The sensing layer composed of p-type rGO thin film with holes as majority carriers. d) An object with positive static charge on its surface attracts the minority carriers in the sensing layer, decreasing the resistance of the sensor. e) Negatively charged surface on the other hand attracts majority carrier holes there by increasing the resistance of the sensor.

**Figure 1a** schematically illustrates the components of flexible proximity sensor. The sensing layer consists of a thin film of rGO obtained via chemical reduction ($N_2H_4$) of vacuum filtrated GO layer. The rGO thin film was then encapsulated between ultrathin polydimethylsiloxane (PDMS) (device thickness = ~18 μm) layers. Sensor fabrication process flow is detailed in **figure S1**.

Synthesis of GO and its chemical reduction into rGO was confirmed by characterization techniques such as transmission electron microscopy, X-ray diffraction, Raman spectra, atomic force microscopy and X-ray photoelectron spectroscopy as explained in **figure S2, S3**. PDMS encapsulation provided mechanical strength and flexibility to the sensor while protecting the rGO layer from moisture. The build of rGO proximity sensor allows it to be worn on skin or on any curvilinear surface as shown in **figure 1b**. The ultrathin nature of the device aids not only high skin conformity, but also promotes better adhesion to skin preventing any obstruction to the movement of subject body.

The sensor response to proximity of an object is partly driven by the triboelectric effect. An electrically neutral object upon contact with another object or atmospheric air (due to friction) experiences charge transfer due to electron movement from one object to another. The retention of charge upon physical separation, rendering one object positively charged and one negatively charged is the triboelectric effect. Hall measurements have shown that the rGO film used in the present work was a p-type material with holes as majority carriers. Owing to the triboelectric effect, an object with positive static charge when brought near the rGO layer, attracts the minority carrier electrons from channel as shown in **figure 1d.** This reduces the recombination rate in the channel leading to higher current in the film, thereby decreasing its resistance. Similarly, an object with negative static charge on its surface attracts the majority carrier holes from rGO channel thus decreasing the current in the device **(figure 1e)**. It can be concluded that the object whose proximity is under observation acts as gate controlling the flow of current in p-channel rGO depending on the polarity of charge, quantity of charge and distance between the charged object and the channel.

In general, field effect transistors (FET) are employed as biosensors by altering the gate terminal with ion sensitive electrode or molecular receptors.[33–36] Adsorption of charged species on gate terminal alters its electric potential thereby changing the current in the FET channel depending on the type of carrier.[37–39] Similarly, when an electrostatically charged object brought close to the sensor as shown in **figure 1**, it acts as gate voltage there by controlling the flow of current in rGO film.

**Performance of rGO sensor in proximity to electrostatic potential:**

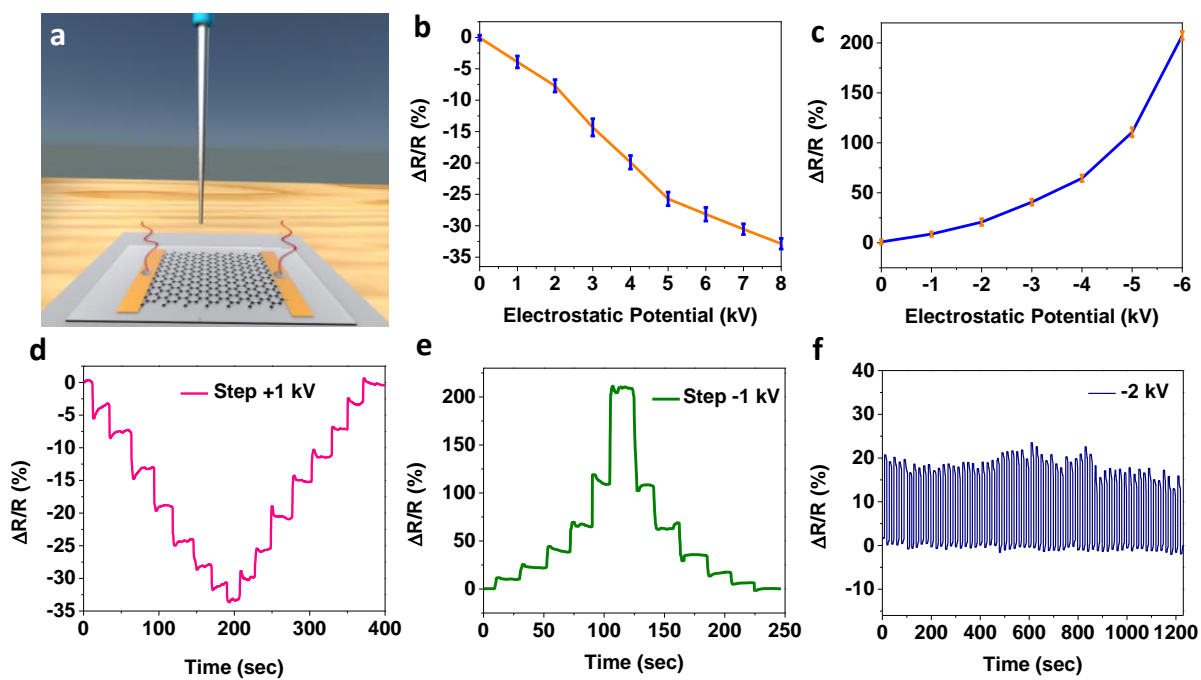

**Figure 2: Response of the sensor to electrostatic potential.** a) Schematic illustrating the set up for the study. The sensor was placed on an electrically neutral aluminum plate and a controlled potential was applied from the pointed top electrode placed at a height of 1 cm from the sensor. b) Decrease in sensor resistance for positive potential from 0 kV to 8 kV. c) Performance of sensor in negative potential from 0 kV to -6 kV. d) Step response of sensor for positive potential with a step height of +1 kV. The sensor exhibited good response with low hysteresis. e) Step response in negative potential for a step height of -1 kV. f) Sensor output was highly repeatable at -2 kV.

To study the response to static electric charges in more detail, the sensor was subjected to a controlled variation of electrostatic potential using electrostatic voltage source. The sensor was placed on an electrically neutral aluminum plate (set to zero electric potential) and specific electrostatic potential (25 kV electrostatic voltage source, at steps of 1 kV) was applied using a point probe (probe tip diameter of 3 mm) placed at a height of 1 cm as shown in **Figure 2a**. The sensor terminals were connected to a source meter (Keithley 2450), controlled via LabVIEW interface, to monitor the resistance of the device at different potentials. The potential was varied in both positive and negative directions and the corresponding normalized change in resistance (ΔR/R %) was plotted (**Figure 2b, c**). With the application of positive potential, sensor recorded decrease in its electrical resistance. As described in the working mechanism, the positive potential attracted the electrons from the p-type rGO allowing more current to flow. With increase in potential, more minority carriers accumulated in the direction of potential and the resistance decreased further. The tests were carried out up to +8 kV, above which the possibility of a dielectric breakdown was high as the electrode gap was small and it would potentially damage the rGO film and the source meter connected to it. Similarly, exposure to negative potential increased the sensor resistance due to accumulation of majority carrier i.e. the holes towards static potential thereby decreasing the current in the film. The sensor exhibited high response with ΔR/R of up to 208% for a negative potential of 6 kV. **Figure 2d, e** show the response of the sensor for a step height of 1 kV. It was observed that the sensor response was higher for negative potential compared to positive potential. Higher sensitivity towards negative potential can be attributed to holes being the majority carriers. Upon exposure to negative potential, the majority carriers get attracted towards the negative field thus reducing their contribution to current conduction in the rGO channel. This is in contrast with the response of the sensor to the positive potential of the same magnitude since attraction of the minority carriers i.e. the electrons towards the positive field does not reflect significantly on the current in the rGO channel. The sensor displayed excellent repeatability when tested at -2 kV (**Figure 2f**). The probe tip was fixed at 1 cm from the sensing layer and the static voltage was turned ON and OFF repeatedly up to 100 cycles. Minor variation in the magnitude of sensor response was observed due to the analog controller of the voltage source and achieving -2 kV precisely and instantly was challenging.

The fabricated sensor was tested to detect proximity of various objects. As the device is sensitive to nearby static charges, the effect of substrate or the body upon which the sensor was mounted had to be investigated. To understand the influence of substrate on the conductivity of rGO, the study was carried out using two methods. First, the sensor was placed on an electrically neutral platform and its response was recorded for proximity of different objects. Later, the flexible sensor was tattooed onto the skin and same experiments were performed.

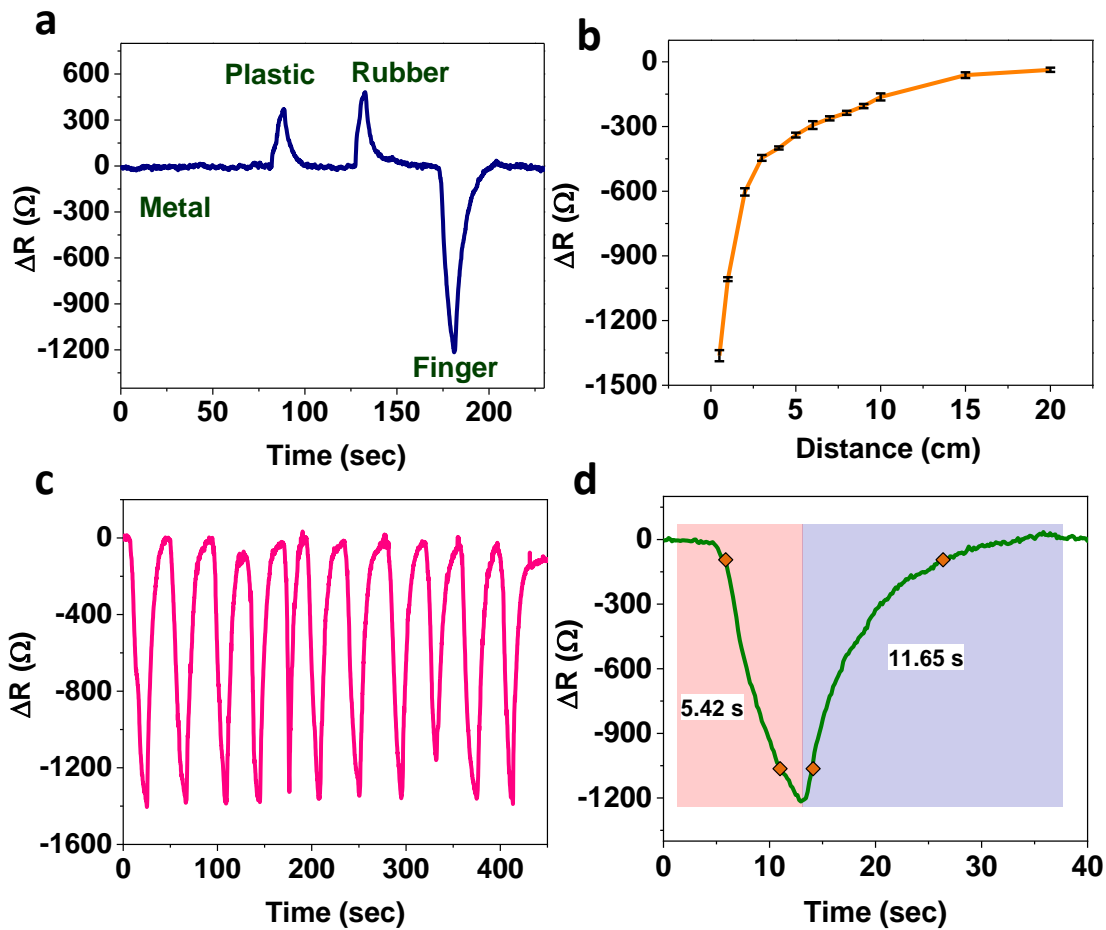

**Figure 3. Sensing proximity of various objects:** a) Sensor response to different objects. Output of sensor depends on the magnitude of static electricity residing on the object, polarity of the charge and distance between sensor to object. b) Proximity of human finger recorded at various distances from the sensor. The sensor could detect the presence of finger up to 20 cm. c) Figure shows repeatability of the device with respect to movement of hand when brought close to sensor up to 5 mm. d) Response time and recovery time of the sensor for the proximity of finger at 0.5 mm distance were 5.42 s and 11.65 s, respectively. Higher recovery time can be attributed to trapping of charge carriers in the interface (rGO-PDMS) defect states which take longer time to de-trap upon removal of charged source.

While testing the sensor placed on a neutral platform, change in electrical resistance was observed within the proximity of an external object. **Figure 3a** shows sensor response when objects such as metal, plastic, rubber and finger were brought close. Metal, being electrically neutral, caused no variations in sensor resistance resulting in a flat resistance curve. Plastic and rubber increased the device resistance due to the presence of negative surface charges.[32] Human finger on the other hand accumulates positive static charges on its surface thereby causing the sensor resistance to decrease. The influence of distance between the object and the sensor on its sensitivity is discussed in detail below.

Response of sensor depends on the magnitude of static charges residing on the object, polarity of the charge and distance between the object and sensor. The change in resistance with respect to distance of the object (in this case, finger) from the sensor was plotted in **Figure 3b**. It can be observed that the sensor resistance registered a significant drop when approached with the human finger. The finger was placed at different heights from the sensor and the corresponding resistance were measured. A large resistance drop was observed as the hand moved close to the sensor. Maximum drop in resistance was obtained when the finger was closest to the sensor i.e. at 5 mm. The sensor could detect the proximity of hand for as far as 20 cm. The **supplementary video S1** demonstrates the sensor performance in action. Continuous monitoring of sensor resistance with hand movement was enabled by LabVIEW. **Figure 3c** indicates the repeatability with respect to finger positioned at 5 mm from the sensor.

Response time and recovery time of the sensor for the proximity of finger (at 0.5 mm) were measured to be 5.42 s and 11.65 s respectively as shown in **Figure 3d**. The observation reveals that the response when the object comes closer is faster than when it moves away. Slow recovery time indicate towards a very interesting phenomenon. As the charged object in proximity is removed, the attracted charge carriers within rGO film go back into the channel and the resistivity is expected to recover. When the charges get attracted to the surface (rGO and PDMS interface), they get trapped in the interface defect states. Upon removal of the charged object, the charge carriers at the interface are unable to go back into the channel immediately as they are trapped in these interface states. The trapped charges slowly de-trap over time and the recovery time is dominated by this effect.

**Wearability of proximity sensor**

It is important to note the ease of wearability afforded by the sensor as the response was recorded upon attaching it on the skin. The device was mounted onto the forearm to demonstrate its proximity sensing characteristics as a wearable sensor. Being ultra-thin, the tattoo sensor was carefully transferred on to the skin in a process similar to that of a temporary tattoo. Transfer process involved the following steps. First, the tattoo sensor on butter paper (refer fabrication process) was placed on the skin with the sensor side facing the skin. The butter paper on top of the sensor was then wetted by sprinkling few drops of water. Wetting of butter paper released the tattoo onto the skin. Finally, the sensor was secured using a transparent sheet of Silhouette transfer sheet to provide better strength and mechanical stability to the contact wires. **Figure 4e** shows the high skin conformability of the tattoo. Previous results while experimenting with the sensor have shown that skin accumulated static charges and induced change in sensor resistance hence, mounting the sensor on skin for wearable application demands the investigation of skin influence on the electrical performance of proximity sensor.

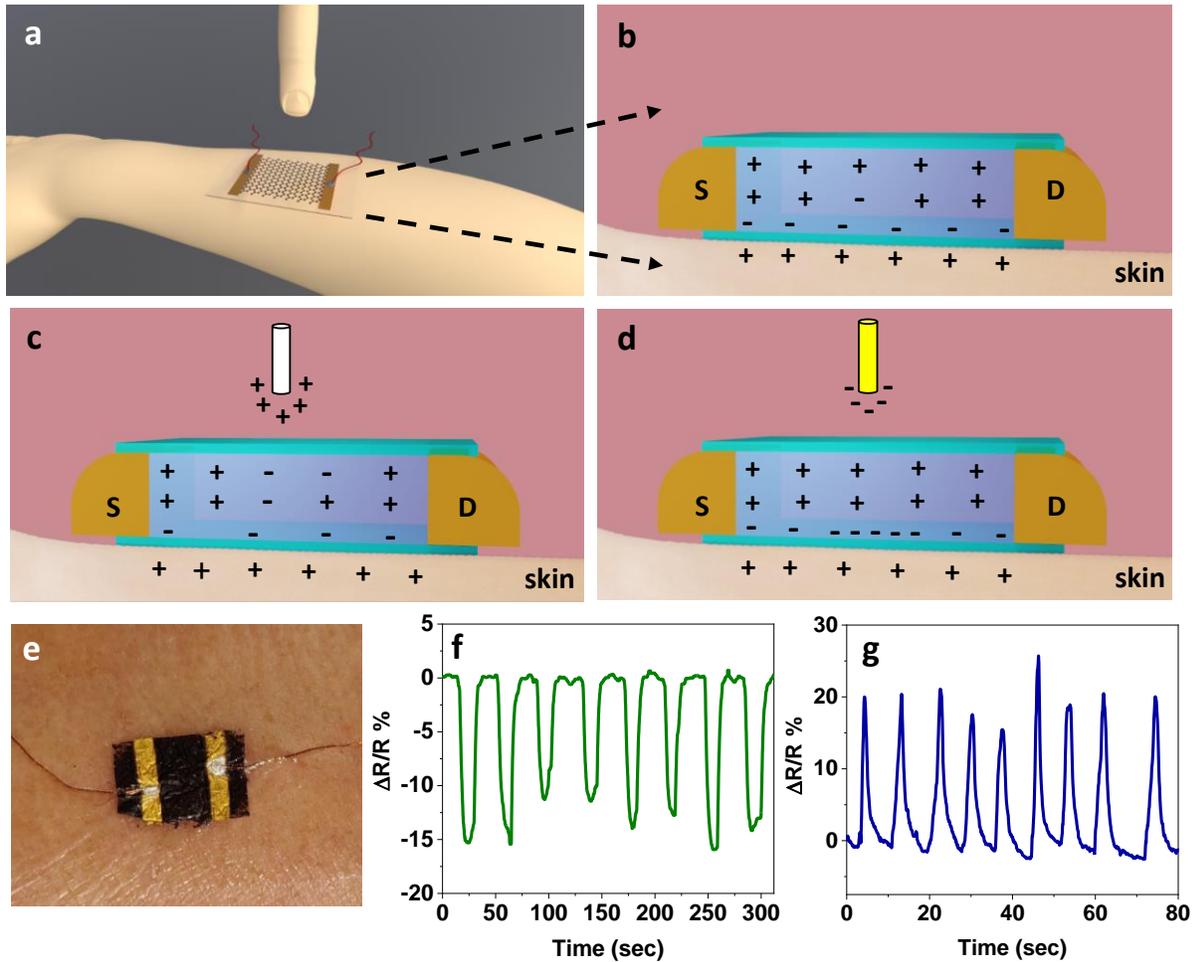

**Figure 4. Proximity sensor tattooed on skin for wearable applications:** a) Schematic of tattoo sensor attached on forearm to detect proximity of human finger. b) The positive static charges on skin forearm attract minority carriers of p-type rGO film causing the sensor resistance to drop. c) Introduction of another object with positive charges residing on its surface reduced the net positive potential experienced on the sensor from both sides, increasing the sensor resistance. d) Negatively charged object increased the net positive potential experienced by the sensor thus decreasing the resistance. e) Photograph of proximity sensor tattooed on the skin and secured via Silhouette paper to demonstrate wearable applications. f) Decrease in sensor resistance for the proximity of a plastic pen placed at ~1 cm from the device. g) Sensor response to proximity of human finger at ~1 cm displayed good repeatability.

**Figure 4a** depicts the schematic of wearable proximity sensor experiment with human finger. The skin, not being electrically neutral, compels the sensor to follow a different working mechanism. The static charges trapped underneath the tattoo sensor cause charge separation in rGO film. The skin is known to harbor positive charges[32] causing the minority carriers i.e. the electrons from the p-type rGO to deviate towards the skin causing the sensor resistance to drop significantly (**Figure 4b**). Upon introduction of an object with positive potential near the sensor attached to skin, the sensor experienced a decrease in the net positive potential from both sides. This lead to decrease in current in the film, increasing its resistance from the previous value (**Figure 4c**). On the other hand, an object with negative static charges increased the net positive charge experienced from skin decreasing the resistance of the sensor as shown in **Figure 4d.** Proximity of finger was recorded as a change in resistance of the sensor as shown in **Figure 4f**. The response of the sensor was found to be reversed in comparison to similar tests conducted earlier by placing the sensor on an electrically neutral platform. Finger proximity to on-body sensor increased its resistance contrary to earlier results when the sensor was placed on neutral platform. Similarly, proximity of plastic pen charged with negative potential placed at ~1 cm from the device decreased the sensor resistance as shown in **Figure 4g.**

**Proximity sensor for human machine interfacing**

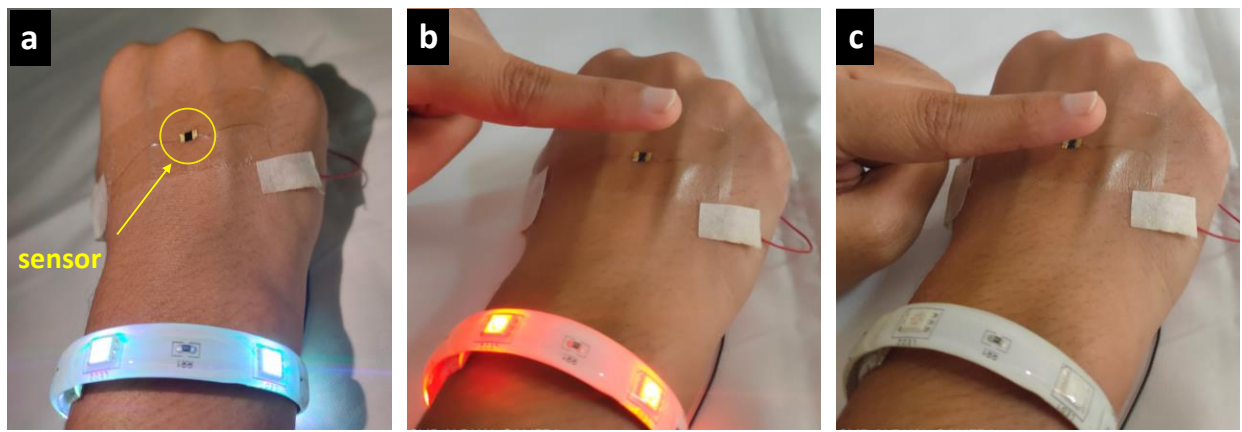

**Figure 5. Demonstration of proximity sensor as a wearable controller:** The wearable proximity sensor was wired to RGB LED bracelet to control its display color and brightness a) Sensor attached to dorsal side of hand is at low resistance state with no finger in proximity. LED displayed blue color in this state. b) An approaching finger increased the sensor resistance, thereby decreasing the controller current and changing the LED color to red. c) Finger located at closest proximity of sensor induced maximum resistance change. LED turns off at the state.

**Figure 5** demonstrates the switching application where the color and intensity control of light emitting diode (LED) was achieved using the wearable proximity sensor. The RGB LED used in this experiment was driven by an Arduino board and the colors were controlled by changing the resistance of the sensor. The ability of the sensor to change its resistance with respect to the proximity of finger was used to control the brightness of individual LED segment (red green and blue). The sensor attached to dorsal side of hand was initially at low resistance state allowing full current through driver circuit, glowing blue LED at maximum intensity (**Figure 5a**). The LED color gradually changed to red with the approach of the finger, corresponding to the function of finger's distance from the sensor (**Figure 5b**). Change in LED color was caused by the static charge residing on the fingertip thereby varying sensor resistance. The driver circuit received minimum current when finger reached closest to the sensor thus turning it OFF **(Figure 5c)**. **Supplementary video S2** demonstrates the control/switching action of wearable proximity sensor. This demonstration showcases the versatility of the sensor in wearable applications especially as an alert mechanism for human and object proximity detection during SARS-CoV-2 pandemic facilitating better social distancing.

The results showcased in this paper highlight how an ultra-thin, highly flexible proximity sensor can be effectively deployed on any kind of surface for short range sensing. **Table 1** gives a comprehensive overview of the important nanomaterial-based proximity sensors reported until now. Most of the reported proximity sensors while capable of detecting either metallic or non-metallic objects did not boast of smooth integration on the surface of the skin for enhanced wearability along with a higher detection range.

**Table 1: Nanomaterial-based proximity sensors and their relative performances**

| Proximity sensor | Sensing mechanism | Objects detected | Maximum sensitivity | Detection limit | Wearability | references |
|---|---|---|---|---|---|---|
| rGO | Electrostatic gating | Non-metallic objects with electrostatic charge on them | $|\Delta R|$ of ~1360 Ω for human finger at WD of 5 mm | Up to 20 cm for human finger | Skin conformal and biocompatible | Present work |
| graphene | Electrostatic shielding and capacity coupling | Non-metallic objects with electrostatic charge on them | $|\Delta I|$ of ~800 µA for rubber rod at WD of 25 mm | >10 cm for an electrostatic source | Fabricated on Si wafer, thick and highly rigid | [32] |
| m-r(CNC/GO) | Proximity of an object distorts electric fields of sensor and causes a change in resistance and capacitance. | only sensitive to human body | ~50% change in relative resistance for finger at WD of 0.2 mm | 6 mm for human finger | Flexible but wearability is not reported | [31] |
| Ni fibers | Capacitive type | Conductive materials only | 1.8% change in potential recorded for palm at WD of 2 mm | 7 cm for palm | Flexible and wearable | [10] |
| Organic crystal (rubrene) | Electrostatic gating | Non-metallic objects with electrostatic charge on them | - | 5 mm for objects | Flexible but wearability is not reported | [11] |
| Ag ink | Capacitive type | Conductive materials only | $|\Delta C|$ of ~0.8 µF for metal plate at WD of 0.1 mm | 10 cm for metal plate | Flexible but wearability is not reported | [7] |
| DPP-DTT based organic transistor | Electrostatic gating | Non-metallic objects with electrostatic charge on them | Max current response of 2.2 µA between WD of 3 mm – 53 mm | ~20 cm for PTFE rod | Fabricated on Si wafer, highly rigid | [12] |
| PTFE | Triboelectric nanogenerator | Only detection of human body is reported | 315 V·m$^{-1}$ | ~11 cm for hand | - | [41] |

*Note: WD = working distance*

**Conclusion:**

The reported work successfully demonstrates flexible proximity sensor with rGO as the sensing layer. The sensor displayed outstanding response to electrostatic potential with high sensitivity which is fundamental in determining the proximity of objects. Detection of objects was possible up to 20 cm which is the largest distance among graphene-based proximity sensors reported till date. Good repeatability of sensor for various objects validates that the sensor doesn't saturate after repeated usage. The skin conformal design with biocompatible packaging allows the sensor to be used for wearable applications. The effect of the skin on the electrical performance of the sensor was analyzed in detail. This study demonstrated a first of its kind utilization of electrostatic gating for proximity sensing of non-metallic entities in a truly wearable fashion. The prevailing world situation as the consequence of highly viral SARs-Cov-2 is increasingly in need of ways and methods to enforce social distancing while transitioning from a state of quarantine to normal activities. A cost effective, wearable, and personalized proximity sensor as reported in this work to alert people against unwarranted touching in public spaces could be instrumental in preventing further spread of the virus.

**Material and Methods:**

**Preparation of GO solution**

GO was synthesized via Modified Hummer's method.[40] A known amount of GO powder was dissolved in DI water and sonicated (SinapTec ultrasonic sonicator, NexTgen Lab 120) to obtain uniform dispersion of 0.5mg/mL GO solution.

**Fabrication:**

A uniform dispersion of 0.5mg/mL GO solution was vacuum filtered on a nitrocellulose filter paper and dried at room temperature for 24 hr. A thin layer (~8 μm) of PDMS was spin-coated on top of GO film at 6000 rpm. PDMS film was dried at 90 °C for 2 hr. The PDMS coated filter paper was then dipped in DMF solution with filter paper face down in order to etch the nitrocellulose paper. Etching process takes about 24 hr. The filter paper free GO layer attached on the PDMS film was fished out rom DMF solution using aluminum foil with PDMS film attached to aluminum and vacuum dried for 12 hr. The GO film was then treated with $N_2H_4$ vapor at 90 °C for 24hr to reduce oxygen functional groups in GO. The rGO film was then dried at 90 °C for 6 hrs. Gold contact pads were sputtered on the film to make electrical connections on which copper wires were attached using silver paste. Finally, a thin layer of PDMS was spin coated at 6000 rpm on the sensing film for protective polymer encapsulation. Bottom aluminum foil was etched in (0.1mg/mL) NaOH solution and rinsed in DI water several times before transferring on the skin for proximity sensing applications.

**Transfer process:**

The aluminum foil attached to the tattoo sensor was etched using (0.1mg/mL) NaOH solution. The free-standing sensor needs a new base to rest on, to avoid the ultrathin PDMS layer crumple on itself. The floating sensor in NaOH solution was fished out using a butter paper, as butter paper assists easy transfer process. The sensor on butter paper was double washed in DI water for 30 min before transferring it on the skin. In order to attach the tattoo sensor on the body, a location on the skin was chosen. The sensor was then placed on the desired location with the tattoo side touching the skin. The butter paper on top was wetted by sprinkling few drops of water. This releases the tattoo onto the skin. Finally, the sensor was secured using a transfer sheet of Silhouette tattoo paper which provides better strength and mechanical stability to the contact wires.

**Disclaimer:**

An informed signed consent was obtained from the subject for wearable sensing experiments.

The author has used self as the subject for wearable experiments. The tattoo sensor was encapsulated in bio compatible polymer (PDMS) material. The Silhouette tattoo paper used to secure sensor is bio compatible. The process of transferring the tattoo onto the skin caused no harm/ irritation to the subject. Hence the approval from a national or institutional ethics board/committee prior to the research was not necessary.

**Acknowledgements**

The authors are grateful to the department of Instrumentation and Applied Physics (IAP) and Centre for Nanoscience and Engineering (CeNSE) at Indian Institute of Science (IISc) for providing necessary facilities to carry out the experiments. The authors acknowledge the support of National Nano Fabrication Facility (NNFC), Micro and Nano Characterization Facility (MNCF) and Packaging lab at the CeNSE. We are thankful to Manu M Pai at IAP and Pavithra B at CeNSE for helping us with the experiments. D.A acknowledges the support of the Office of Naval Research (ONR).

**TOC Image**

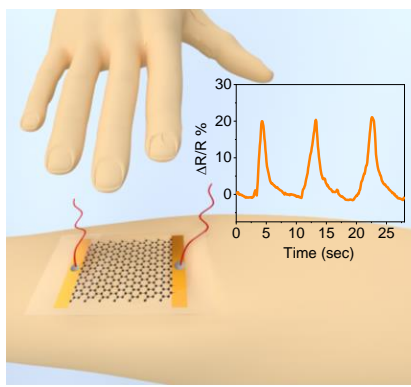